\documentclass[conference]{IEEEtran}
\IEEEoverridecommandlockouts
\usepackage[utf8]{inputenc}
\usepackage{cite}
\usepackage{amsmath,amssymb,amsfonts}
\usepackage{algorithm}
\usepackage{algorithmic}
\usepackage{graphicx}
\usepackage{caption}
\usepackage{subcaption}
\usepackage{textcomp}
\usepackage{xcolor}
\def\BibTeX{{\rm B\kern-.05em{\sc i\kern-.025em b}\kern-.08em
    T\kern-.1667em\lower.7ex\hbox{E}\kern-.125emX}}

\usepackage[T1]{fontenc}
\usepackage{newtxtext,newtxmath}

\usepackage[skip=3pt]{caption}

\usepackage{enumitem}
\setitemize{topsep=2pt,parsep=0pt,partopsep=0pt,leftmargin=15pt}

\usepackage[inline]{./trackchanges}%
\addeditor{Shun}
\addeditor{Qing}
\addeditor{Changheng}

\usepackage{xspace}
\newcommand{\SysName}{\mbox{ScreenAnt}\xspace}

\begin{document}
\title{ScreenAnt: Transparent On-Screen Antennas for 6G
\thanks{This work is supported by the 6thSense project, which has received funding from the European Union’s Framework Programme for Research and Innovation under the HORIZON-MSCA-DN-2022 Grant Agreement No.~101119652.}
}

\author{\IEEEauthorblockN{Shun Zhuge and Qing Wang}
\IEEEauthorblockA{Delft University of Technology, The Netherlands \\
Emails:\{s.zhuge, qing.wang\}@tudelft.nl}
}

\maketitle

\begin{abstract}

6G will require on-device antenna systems to operate at ultra-high frequency bands, achieve robust beamforming on the compact user devices, and be blockage-robust. Conventional \textit{edge-mounted} antennas on devices have limited apertures, suffer from the \textit{`death grip'} caused by user-induced blockage, and have poor scalability at mmWave and sub-THz bands. To address these issues, motivated by the rapid evolution of transparent~materials and antennas, we propose \textit{ScreenAnt} in this work---which integrates a transparent antenna array onto the screens of future mobile devices. Specifically, we propose using a transparent \textit{on-screen}~uniform~planar array and develop a framework to model its electromagnetic property, spatial configuration, and blockage robustness under realistic user-induced blockage. We also design a gradient-ascent-based algorithm to efficiently optimize power and phase control of on-screen antennas to maximize ScreenAnt's spectral efficiency. Our thorough simulations show that the proposed ScreenAnt can increase the uplink spectral efficiency by over 50\% compared to edge-mounted antennas at 28 GHz, and by more than 150\% at 300 GHz. ScreenAnt also demonstrates strong robustness against user-induced blockage, paving the way for practical and high-capacity 6G user device designs.

\end{abstract}

\section{introduction}
\label{introduction}

Each evolution in cellular networks reshaped both~the~wireless capacity and the antenna form factor of user devices.~Early 1G phones featured 
exposed whip antennas supporting only analog voice, whereas 4G and 5G smartphones place multiple compact antennas \textit{on the edges of smartphones} to support~the multiple-input multiple-output (MIMO)~\cite{Wang2024Antenna} for broadband connectivity. 
As we look toward 6G, a pressing challenge arises: \textit{within~the constrained space expected on future 6G mobile devices, how should their antenna systems evolve to operate at increasingly higher radio frequency bands?} 

6G is expected to use millimeter-wave (mmWave) and sub-THz bands to achieve extremely high data rates and to have the joint communication and sensing (JCAS) capability~\cite{liu2022isac}. These frequencies and capability demand dense and high-gain antenna arrays capable of high-resolution beamforming and robustness against blockage caused by touch interactions of users with their mobile devices~\cite{raghavan2021hand}. Achieving these demands within the compact form factors of mobile devices, such as smartphones and smartwatches, poses design challenges~\cite{mertens2024antennas}.

\begin{figure}[!t]
    \centering
\includegraphics[width=.75\columnwidth]{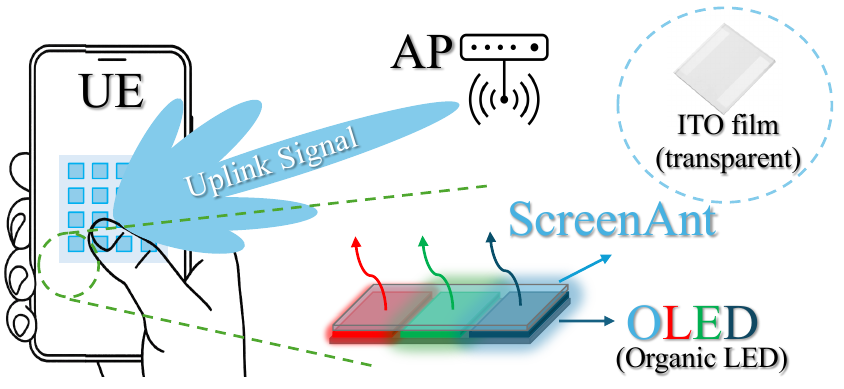}
    \caption{The illustration of our proposed {\textit{\SysName}} that~integrates transparent on-screen antenna array onto future 6G UEs.} 
    \vspace{-6mm}
    \label{fig_system_ill}
\end{figure}

On 4G/5G user equipment (UE), the antenna integration is constrained by metallic enclosures and limited device geometry. Commercial 4G/5G smartphones employ edge-mounted antenna elements through chassis cutouts, distributing single-element modules along the device frame to improve the spatial coverage and isolation~\cite{yuan_wideband_2021,jaglan_10_2021,wang_compact_2020}. While these solutions simplify mechanical design, they inherently limit the available aperture and remain highly susceptible to user-induced blockage—commonly known as \textit{death~grip}~\cite{raghavan2021hand}. These structural limitations are 
becoming critical bottlenecks for higher-frequency operation, motivating new antenna deployment strategies that can scale effectively with the performance demands of 6G.

In the past few years, we have witnessed the rise~of~research and development in \textit{transparent antennas} in both academia and industry, driven by transparent antennas' unique capability to seamlessly integrate antenna elements into visually critical surfaces, such as windows and device screens~\cite{silva2020, liu_enhancing_2023, inotec2020, sayem2022, sun2023, liu2024}.
Leveraging transparent conductive materials like \textit{indium tin oxide (ITO)}, \textit{silver nanowires}, or \textit{graphene}, these transparent antennas offer significant benefits such as improved aesthetics and minimized form-factor constraints. 
In parallel, several antenna-on-display (AOD) concepts have been proposed, aiming to relocate antennas beneath or around the display to free chassis space and improve radiation efficiency~\cite{park2019aod,hong_mmwave_2020,chou2025smartphone}.
While both approaches represent meaningful advances toward on-screen wireless architectures, their practical 
designs remain limited. Most AOD designs position metallic elements along the bezel and focus only on electromagnetic performance, whereas transparent antenna studies mainly consider isolated single elements or small subarrays for aesthetic use~\cite{hassan2023transparent}. Neither approach has addressed large-scale transparent array integration or its system-level modeling, such as blockage modeling, spatial configuration, and optimization under real device constraints. These gaps motivate a transparent and screen-integrated scalable antenna array architecture that can simultaneously enhance communication efficiency, sensing capability, and robustness to user-induced blockage on 6G UEs. 

In this paper, we propose a novel \textit{transparent on-screen~antenna array architecture} for 6G UEs, referred to as \textit{\SysName}, as illustrated in Fig.~\ref{fig_system_ill}. Our {\SysName} has two innovations:  
(i) relocation of antenna arrays from device edges to the screen surface, which maximizes effective aperture and minimizes hand blockage impact;  
(ii) utilization of transparent conductive materials (e.g., ITO films) that enable the screen to transmit wireless signals while preserving the visual quality of the display.  
We summarize the main contributions below: 
\begin{enumerate}
    \item We propose and analyze a novel on-screen antenna architecture for future compact 6G UEs: a fully transparent uniform planar array integrated onto the screens of UEs.
    
    \item We develop a unified modeling framework that captures the spatial, electromagnetic, and blockage-aware characteristics of the proposed on-screen antenna architecture, enabling system-level analysis under dynamic conditions.
    
    \item We design a structure-aware optimization algorithm to maximize the spectral efficiency (SE) of \SysName, considering device-side constraints. Our thorough simulation results demonstrate the significant gains of \SysName in uplink SE and blockage robustness compared to conventional edge-mounted antenna designs, achieving over 50\% higher SE at 28 GHz and 150\% improvement at 300 GHz.
\end{enumerate}


\section{Modeling the On-Screen Antenna Array}
\label{sec_model}

\subsection{System Model}
Fig.~\ref{fig_system_ill} illustrates the architecture of our proposed {\SysName}. The proposed {\SysName} enables communication between the UE and the access point (AP) through the on-screen antenna array.
Physically, the transparent antennas of {\SysName} are implemented using ITO–based transparent conductive films embedded on the front glass of the UE's screen, functionally replacing conventional edge-mounted antennas. The high optical transparency of ITO enables the organic light-emitting diode (OLED) panel to render visual content without noticeable degradation, while the same conductive layer simultaneously serves as the radiating medium for wireless signal transmission. Compared with the narrow device edge areas, the much larger screen area allows the deployment of more antenna elements in an integrated array rather than being individually distributed along the edges.
This configuration enables more precise beamforming through dynamic adjustment of the transmit power and phase of each antenna element. Moreover, since users naturally avoid covering the display area during operation, the proposed on-screen placement inherently reduces hand blockage.

\begin{figure}[!t]
    \centering
    \includegraphics[width=.6\columnwidth]{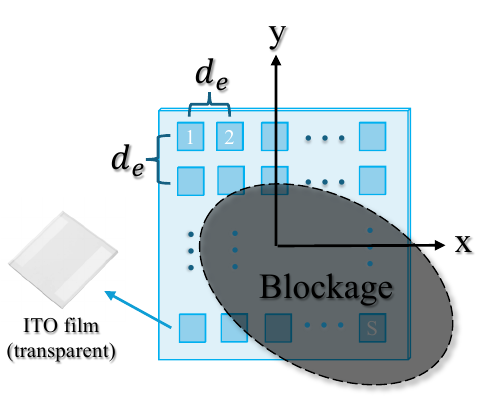}
    \vspace{-2mm}
    \caption{Layout of the transparent on-screen antenna array.}
    \label{fig_system_layout}
    \vspace{-6mm}
\end{figure}

The detailed layout of the transparent on-screen antenna array of {\SysName} is shown in Fig.~\ref{fig_system_layout}, where the transparent antenna array is centrally positioned on the UE's screen with its geometric center located at $(0_x, 0_y)$.
Let $\mathcal{S} = \{1, 2, \cdots, S\}$ denote the set of antenna elements, where $S = S_x \times S_y$, and $S_x$ and $S_y$ represent the number of antenna elements along the $x$- and $y$-axes, respectively.
The array is uniformly formed (i.e., $S_x=S_y$), and the element spacing is denoted by $d_e$.

The Cartesian coordinates of the $s$-th transparent on-screen antenna element, denoted as $(s_x, s_y)$, are given by
\begin{align}
s_x = d_e \left(-\frac{S_x - 1}{2} + \overline{s}_x \right), \ \
s_y = d_e \left(\frac{S_y - 1}{2} - \overline{s}_y \right),
\end{align}
where $s \in \mathcal{S}$, $\overline{s}_x = \mathrm{mod}(s - 1, S_x) + 1$, and $\overline{s}_y = \left\lfloor \frac{s - 1}{S_x} \right\rfloor + 1$, with $\mathrm{mod}(\cdot,\cdot)$ and $\lfloor \cdot \rfloor$ denoting the modulo and the floor operations, respectively.
The Euclidean distance between two antenna elements $s$ and $s' \in \mathcal{S}$ is derived as follows
\begin{equation}
d_{s,s'} = \sqrt{(s_x - s'_x)^2 + (s_y - s'_y)^2}.
\label{d_ss}
\end{equation}
Each on-screen antenna element $s$ is associated with a complex phase coefficient $\phi_s = e^{j\theta_s}$, where $\theta_s \in (0, 2\pi]$ denotes the phase shift, and a transmit power allocation $p_s \ge 0$. The resulting beamforming weight for $s$ is expressed as
\begin{equation}
w_s = \sqrt{p_s}\phi_s.
\label{w}
\end{equation}
By stacking all elements, the overall beamforming vector 
is
\begin{equation}
\mathbf{W} = [w_1, w_2, \cdots, w_S]^T \in \mathbb{C}^{S \times 1}.
\end{equation}

\subsection{Channel Model}
We assume a flat Rayleigh fading channel. The equivalent uplink channel from the transparent on-screen antenna  array of the UE to the AP is modeled as
\begin{equation}
\mathbf{H}_0 \sim \mathcal{CN} \left( \mathbf{0},
\alpha \cdot \ell(d_\text{AP}) \cdot \mathbf{R}_c \right),
\label{eq_channel}
\end{equation}
where $\ell(d_{\text{AP}})$ represents the distance-dependent path loss, and $\mathbf{R}_c \in \mathbb{C}^{S \times S}$ denotes the spatial correlation matrix among antenna elements. To account for the conductivity loss of the transparent materials such as ITO, a transmission efficiency factor $\alpha \in (0,1]$ is introduced, representing the ratio of the radiated power of antennas made from transparent conductive materials to that of conventional metal-based antennas.

The large-scale path loss is modeled as
\begin{equation}
\ell(d_\text{AP}) = \ell(d_0) \left( \frac{d_\text{AP}}{d_0} \right)^{-\eta},
\quad
\ell(d_0) = \left( \frac{\lambda}{4\pi d_0} \right)^2,
\label{eq_pl}
\end{equation}
with the reference distance $d_0 = 1~\mathrm{m}$ and path loss exponent $\eta > 2$, where $\lambda = c / f_0$ denotes the carrier wavelength corresponding to carrier frequency $f_0$ and speed of light $c$.

Assuming an isotropic scattering environment, the spatial correlation matrix is expressed as
\begin{equation}
[\mathbf{R}_c]_{s,s'} = \mathrm{sinc}\left( \frac{2 d_{s,s'}}{\lambda} \right),
\quad \forall s,s' \in \mathcal{S},
\end{equation}
where $d_{s,s'}$ is the Euclidean distance between the on-screen antenna elements $s$ and $s'$, as previously defined in~\eqref{d_ss}.

\subsection{Blockage Model}
\label{subsec_blockage}
Based on the channel model in~\eqref{eq_channel}, the effect of user-induced blockage is further incorporated as follows.
Although the {\SysName} design greatly mitigates the likelihood of blockage caused by the user’s hand, interaction between the fingers and the screen may still lead to temporary signal attenuation, as illustrated by the shadowed region in Fig.~\ref{fig_system_layout}, which indicates the blocked area of the array.
To capture this effect, a blockage mask $\mathbf{B}\in\{0,1\}^{S\times1}$ is introduced, where $B_s=1$ denotes that antenna element $s$ is blocked, and $B_s=0$ otherwise. The post-blockage channel state information (CSI) is modeled as
\begin{equation}
\mathbf{H}_\text{blk} = \mathbf{H}_0 - (1-\beta)\big(\mathbf{B}\odot\mathbf{H}_{0}\big),
\label{eq_blocked_channel}
\end{equation}
where $\odot$ denotes the Hadamard (element-wise) product, and $\beta\in(0,1]$ represents the effective attenuation factor summarizing the average overlap between the finger-contact region and the active on-screen aperture during typical interactions.
This mask-based abstraction provides a unified and reproducible framework for evaluating structural robustness by decoupling user-behavioral randomness from array design.

\paragraph*{Remark on dynamic blockage}
In practice, the blockage effect can vary over time due to changes in touch trajectory and contact area. A more refined model can allow both the blockage pattern and attenuation to evolve dynamically, i.e.,
\begin{equation}
\mathbf{H}_\text{blk}(t)
= \mathbf{H}_0 - \big(\mathbf{I}-\mathbf{K}(t)\big)\big(\mathbf{B}(t)\odot\mathbf{H}_{0}\big),
\label{eq_block}
\end{equation}
where $\mathbf{B}(t)$ captures the time-varying blocked antenna elements, and $\mathbf{K}(t)=\mathrm{diag}(\boldsymbol{\beta}(t))$ represents the time-varying attenuation matrix with $\boldsymbol{\beta}(t) = [\beta_1(t), \beta_2(t), \cdots, \beta_S(t)]^T$.
Each $\beta_s(t)\in(0,1]$ quantifies the instantaneous attenuation of element $s$ under dynamic touch conditions.

Extending the analysis to stochastic and time-varying blockage processes is an interesting direction. The architecture of {\SysName} is compatible with JCAS-based sensing of such dynamics~\cite{liu2022isac,meng2023vehicular}, enabling adaptive optimization based on real-time channel information, although a dedicated sensing design is beyond the scope of this work and left for future study.
\section{Optimization Framework}
\label{sec_analysis}

Based on the developed system, channel, and blockage~models, we now present the optimization framework we design for the uplink SE maximization in our {\SysName} architecture. 

\subsection{Problem Formulation}
Let $t \sim \mathcal{CN}(0,1)$ denote the transmitted symbol with $\mathbb{E}[|t|^2]=1$. The received baseband signal at the AP is:
\begin{equation}
    y = \mathbf{H}_\text{blk} \mathbf{W} t + n,
\end{equation}
where $\mathbf{W} \in \mathbb{C}^{S \times 1}$ is the beamforming vector defined in \eqref{w}, and $n \sim \mathcal{CN}(0,\sigma^2)$ denotes the additive white Gaussian noise.

The corresponding uplink signal-to-noise ratio (SNR) at the AP is derived as
\begin{equation}
    \gamma = \frac{|\mathbf{H}_\text{blk} \mathbf{W}|^2}{\sigma^2}.
\end{equation}

To maximize the uplink SE, we jointly optimize the transmit power and phase of each antenna element. We assume that the CSI is perfectly known at the UE to characterize the theoretical performance upper bound, which can be approximately obtained through uplink training and channel estimation \cite{Sattari2024}.

The optimization problem is formulated as
\begin{align}
    \max_{p_s,\phi_s}\quad & \text{SE}=\log_2\left(1+\gamma\right) \label{SE}\tag{12a}\\
    \text{s.t.}\quad&\sum^S_{s=1}p_s\leq P,\tag{12b}\\
    & p_s\geq 0, \quad \forall s \in \mathcal{S}, \tag{12c} \\
    & |\phi_s|=1, \quad \forall s \in \mathcal{S} \label{phi_s}\tag{12d},
\end{align}
\setcounter{equation}{12}where $\phi_s = e^{j\theta_s}$ denotes the phase coefficient of the on-screen antenna element $s$ with phase angle $\theta_s \in (0, 2\pi]$, $p_s$ represents the transmit power allocated to the antenna element $s$, and $P$ denotes the total available transmit power at the UE.

\subsection{Uplink SE Optimization}

To solve this problem \eqref{SE} efficiently, we design an iterative gradient ascent algorithm that updates both phase shifts and power allocation. The constant-modulus constraint in \eqref{phi_s} is enforced throughout by formulating the gradient in terms of the phase angle $\theta_s$. The algorithm proceeds as follows:

\vspace{1mm}
\textit{Step 1: Initialization.}
Initialize the phase shifts $\theta_s$ and power allocations $p_s$ using a uniform random distribution. To mitigate the risk of convergence to a local optimum, multiple random initializations are generated, and the one yielding the highest initial $\text{SE}$ is selected for subsequent optimization. 

\vspace{1mm}
\textit{Step 2: Gradient computation.} 
For each one-screen antenna element $s \in \mathcal{S}$, compute the partial derivative of the achievable 
$\text{SE}$ with respect to the phase shift $\theta_s$: 
\begin{align}
    \frac{\partial{\text{SE}}}{\partial{\theta_s}}&=\frac{\partial{\text{SE}}}{\partial{\gamma}}\frac{\partial{\gamma}}{\partial{\theta_s}} \label{pd}\\
    &=\frac{1}{\ln{2}\left(1+\gamma\right)}\frac{\partial{\frac{|\mathbf{H}_\text{blk} \mathbf{W}|^2}{\sigma^2}}}{\partial{\theta_s}} \notag\\
    &=-\frac{2\sqrt{p_s}}{\ln(2)(1 + \gamma)\sigma^2} \Im \left\{  [\mathbf{H}_\text{blk}]_se^{j \theta_s}\left( \mathbf{H}_\text{blk} \mathbf{W} \right)^H \right\}. \notag
\end{align}

Similarly, compute the derivative of the achievable $\text{SE}$ with respect to the allocated transmit power $p_s$:
\begin{align}
    \frac{\partial{\text{SE}}}{\partial{p_s}}&=\frac{\partial{\text{SE}}}{\partial{\gamma}}\frac{\partial{\gamma}}{\partial{p_s}}\label{pd_p}\\
    &=\frac{1}{\ln{2}\left(1+\gamma\right)}\frac{\partial{\frac{|\mathbf{H}_\text{blk} \mathbf{W}|^2}{\sigma^2}}}{\partial{p_s}} \notag\\
    &=\frac{1}{2\ln(2)(1 + \gamma)\sigma^2\sqrt{p_s}} \Re \left\{[\mathbf{H}_\text{blk}]_s\phi_s \left( \mathbf{H}_\text{blk} \mathbf{W} \right)^H \right\}. \notag
\end{align}

\vspace{1mm}
\textit{Step 3: Gradient normalization.} 
To ensure stable updates, we normalize the calculated gradients as follows:
\begin{align}
    \frac{\partial{\text{SE}}}{\partial{\theta_s}} \leftarrow \frac{2\pi}{\rho}\frac{\partial{\text{SE}}}{\partial{\theta_s}},   \quad 
    \frac{\partial{\text{SE}}}{\partial{p_s}} \leftarrow \frac{P}{\delta}\frac{\partial{\text{SE}}}{\partial{p_s}},  \label{n_p}
\end{align}
where $\rho = \max_{s \in \mathcal{S}} \left| \frac{\partial{\text{SE}}}{\partial{\theta_s}} \right|$ and $\delta = \max_{s \in \mathcal{S}} \left| \frac{\partial{\text{SE}}}{\partial{p_s}} \right|$.

\vspace{1mm}
\textit{Step 4: Update.}
Finally,  we update the phase shifts and the power allocations using gradient ascent as follows:
\begin{align}
    \theta_s \leftarrow \theta_s+\mu\frac{\partial{\text{SE}}}{\partial{\theta_s}}, \ \  
    p_s \leftarrow p_s+\mu\frac{\partial{\text{SE}}}{\partial{p_s}}, \ \forall s \in \mathcal{S} \label{u_p},
\end{align}
where $\mu > 0$ is the Armijo step size determined via backtracking line search. 

\vspace{1mm}
The {\textit{steps 2--4}} are repeated until the fractional increase of $\text{SE}$ falls below a predefined threshold or the maximum number of iterations is reached. The complete procedure of the proposed algorithm is summarized in Algorithm~\ref{alg_1}.

\begin{algorithm}[t]
\caption{Gradient Ascent Algorithm for Solving \eqref{SE}}
\label{alg_1}
\begin{algorithmic}[1]
\STATE \textbf{Input:} $\textbf{H}_\text{blk}$ and $\textbf{W}$
\STATE Initialize $\theta_s$ and $p_s$
\STATE \textbf{REPEAT}
\STATE \hspace{1em} Calculate the partial derivatives of $\text{SE}$ with respect to \\ \hspace{3.5mm} $\theta_s$ and $p_s$ using \eqref{pd} and \eqref{pd_p}
\STATE \hspace{1em} Normalize the partial derivatives using \eqref{n_p}
\STATE \hspace{1em} Update $\theta_s$ and $p_s$ using \eqref{u_p}
\STATE \hspace{1em} Calculate the uplink 
$\text{SE}$ by applying \eqref{SE}
\STATE \textbf{UNTIL} The fractional increase in $\text{SE}$ is below a preset threshold or the maximum number of iterations is reached
\STATE \textbf{OUTPUT:} $\theta_s$ and $p_s$
\end{algorithmic}
\end{algorithm}
\section{Performance Evaluation}
\label{sec_eval}

In this section, we evaluate the 
performance of~{\SysName}. 

\textbf{Benchmark.}
To provide a comprehensive comparison, the conventional edge-mounted antenna design, referred to as \textit{EdgeAnt}, is adopted as the benchmark.
For a fair comparison, EdgeAnt employs the same number of antenna elements and the same total transmit power as our {\SysName}, with equal per-antenna-element power and no phase control, as achieving coherent phase alignment across distributed edge antennas is impractical on compact devices.
Although such a dense edge-mounted layout is unrealistic for UEs, 
this non-coherent configuration provides a conservative lower-bound reference to assess the performance gain of {\SysName} under identical hardware constraints.
For completeness, an ideal coherent EdgeAnt could serve as an upper bound; however, this case is not considered here due to the infeasibility of maintaining full phase coherence in distributed edge-mounted architectures.
In addition, all antenna elements of EdgeAnt are assumed to be conventional metallic units with unity radiation efficiency. 

\textbf{Setup.}
Unless otherwise specified, the following parameters are adopted for all simulations. 
The UE is placed at a fixed distance of $d_\text{AP}=3$~m from a single-antenna AP with a gain of 0~dBi. 
The carrier frequency is $f_0=28$~GHz, corresponding to a wavelength of $\lambda \approx 10.7$~mm. 
As modeled in \eqref{eq_channel}, the wireless channel follows a flat Rayleigh fading model with a reference distance $d_0=1$~m and a path-loss exponent $\eta=2.5$. 
The receiver noise power is $\sigma^2=-110$~dBm, and the total transmit power at the UE is $P=23$~dBm. 

For our 
ScreenAnt, the on-screen antenna element spacing is $d_e=0.5\lambda$, ensuring half-wavelength separation to avoid grating lobes while maintaining compact screen integration. 
The default array size is configured as $S_x \times S_y = 7 \times 7$, yielding a total of $S=49$ transparent antenna elements. 
A transparency efficiency factor of $\alpha=0.85$ is applied to model the conductivity reduction of transparent materials (e.g., ITO). 

For blockage scenarios, a default blockage attenuation coefficient of $\beta=0.1$ is used. The default blockage ratio is set to 50\%, implying that half of the array elements are blocked. 
In the proposed optimization framework shown in Algorithm~\ref{alg_1}, ten random initializations are used, and the maximum number of iterations is limited to five, as the convergence is typically achieved within two iterations. 
All the simulation results are averaged over $10^5$ independent Monte Carlo realizations.

\subsection{Evaluations under Different System Parameters}


We first evaluate {\SysName}'s uplink SE performance under non-blockage conditions
, to evaluate the impact of the transparency efficiency factor~$\alpha$, the number of antenna elements~$S$, the total transmit power $P$, the distance between UE and AP $d_{\text{AP}}$, and the carrier frequency $f_0$ on the system performance.

\begin{figure}[!t]
  \centering
  \begin{minipage}[t]{0.48\columnwidth}
    \centering
    \includegraphics[width=\linewidth]{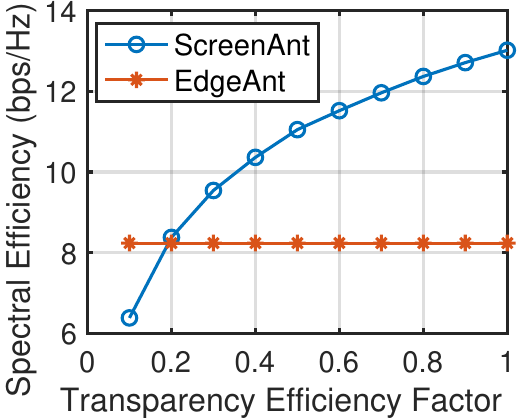}
    \captionof{figure}{SE versus transparency efficiency factor $\alpha$.} 
    \label{fig_transparency}
  \end{minipage}\hfill 
  \begin{minipage}[t]{0.48\columnwidth}
    \centering
    \includegraphics[width=\linewidth]{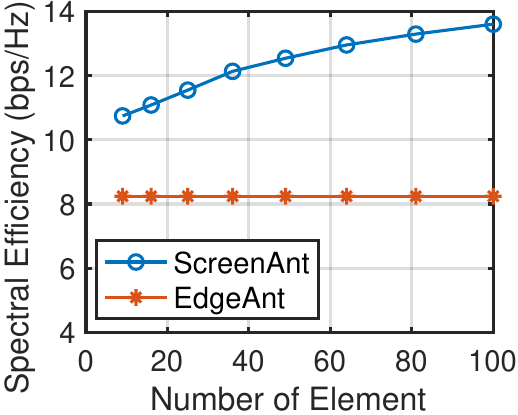}
    \captionof{figure}{SE versus the number of antenna elements \textit{S}.} 
    \label{fig_size} 
  \end{minipage}
  \vspace{-6mm}
\end{figure}

\textbf{SE vs. the antenna's transparency efficiency factor.}
As a key distinction between transparent and conventional metallic antennas, we first examine the effect of the transparency efficiency factor~$\alpha$, which characterizes the tradeoff between optical transparency and electromagnetic conductivity of the antenna material.
As illustrated in Fig.~\ref{fig_transparency}, the achievable SE of {\SysName} increases monotonically with~$\alpha$ but with a gradually diminishing slope, rising from approximately 6~bps/Hz at $\alpha=0.1$ to 13~bps/Hz at $\alpha=1$, since higher transparency efficiency implies improved electrical conductivity, leading to enhanced antenna performance.
The achievable SE of EdgeAnt remains nearly constant at around 8~bps/Hz, as it employs conventional opaque antennas that are unaffected by~$\alpha$.
In practice, materials with lower~$\alpha$ typically exhibit higher optical transparency but poorer electromagnetic conductivity, leading to a tradeoff between communication performance and visual quality.
These results show that {\SysName} can still achieve a clear SE advantage over EdgeAnt with moderately transparent materials ($\alpha=0.2$), confirming the feasibility of balancing communication efficiency and content displaying in reality. 

\textbf{SE vs. the number of antenna elements.}
We next examine the effect of the number of antenna elements.
As shown in Fig.~\ref{fig_size}, the achievable SE of {\SysName} increases with the number of antenna elements due to the enlarged effective aperture and enhanced beamforming gain.
In contrast, the achievable SE of EdgeAnt remains nearly constant at around 8~bps/Hz, as its non-coherent transmission provides no array gain regardless of the number of antenna elements.
This result highlights the limitation of edge-mounted designs, where the difficulty of achieving coherent modulation prevents significant performance improvement even when a large number of antennas are deployed along the device edges.
In practice, the array size of {\SysName} is constrained by the screen dimensions and the physical size of the antenna elements.
Nevertheless, even with a compact array configuration, {\SysName} still achieves a notable SE gain over EdgeAnt, demonstrating the clear advantage of on-screen array integration. 
Considering the higher operating frequencies envisioned for 6G, the antenna element size will further decrease, making the deployment of larger on-screen arrays increasingly feasible.
Overall, {\SysName} achieves an SE improvement of approximately 3~bps/Hz as the number of antenna elements increases from 16 to 100, confirming the benefit of an integrated on-screen planar array.

\begin{figure}[!t]
  \centering
  \begin{minipage}[t]{0.48\columnwidth}
    \centering
    \includegraphics[width=\linewidth]{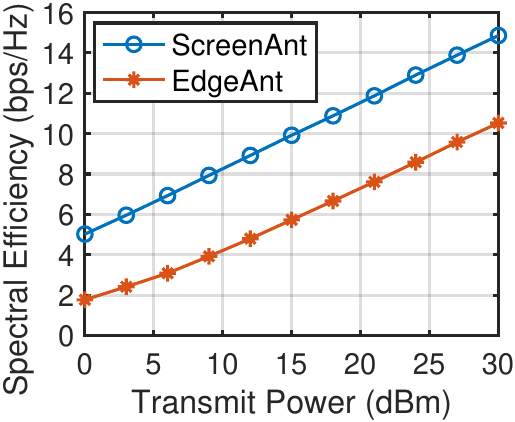}
    \captionof{figure}{SE versus the transmit power $P$.} 
    \label{fig_power}
  \end{minipage}\hfill
  \begin{minipage}[t]{0.48\columnwidth}
    \centering
    \includegraphics[width=\linewidth]{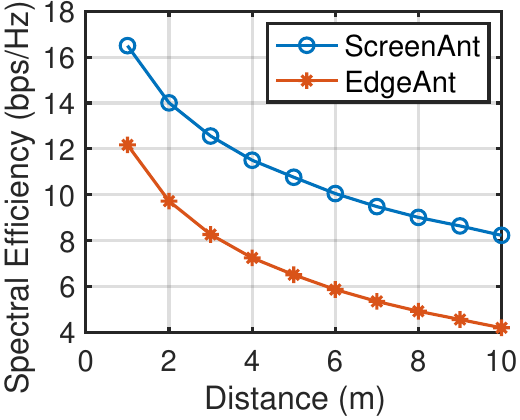}
    \captionof{figure}{SE versus the distance between UE and AP $d_\text{AP}$.} 
    \label{fig_distance}
  \end{minipage}
  \vspace{-3mm}
\end{figure}

\textbf{SE vs. total antenna power budget.}
We continue to evaluate the achievable SE when the UE transmits with different total power $P_t$. The results are shown in Fig.~\ref{fig_power}.
As expected, the SE under both ScreenAnt and EdgeAnt increases almost linearly with the transmit power budget. However, across the entire power range, ScreenAnt consistently outperforms EdgeAnt due to ScreenAnt's larger effective aperture and coherent beamforming capability.

\textbf{SE vs. distance between the UE and the AP.}
Fig.~\ref{fig_distance} shows the achievable SE as a function of $d_\text{AP}$, the distance between the UE and the AP.
As expected, the SE of both ScreenAnt and EdgeAnt decreases with increasing distance due to path loss.
Nevertheless, ScreenAnt consistently maintains higher SE across all distances because of its stronger array gain and more stable beamforming capability.
The performance gain of {\SysName} becomes more evident at larger~$d_\text{AP}$, indicating that the transparent on-screen array of {\SysName} achieves better long-range communication efficiency compared with EdgeAnt.

\begin{figure}[!t]
    \centering
    \includegraphics[width=0.99\columnwidth]{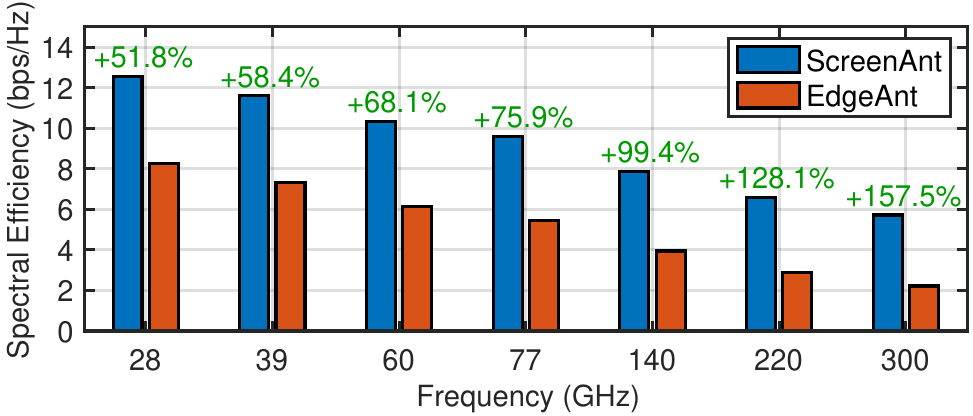}
    \captionof{figure}{SE versus carrier frequency $f_0$ (without blockage).} 
    \label{fig_frequency}
    \vspace{-4mm}
\end{figure}

\textbf{SE vs. carrier frequency.}
We further investigate the impact of carrier frequency on the uplink SE performance.
As shown in Fig.~\ref{fig_frequency}, the achievable SE of both {\SysName} and EdgeAnt decreases as the carrier frequency increases from 28 GHz to 300 GHz, mainly due to the higher path loss at shorter wavelengths.
Nevertheless, {\SysName} experiences smaller degradation, and its relative SE gain over EdgeAnt increases with frequency, from approximately 52\% at 28 GHz to over 150\% at 300 GHz.
This trend demonstrates that the on-screen array architecture exhibits stronger resilience to high-frequency propagation loss, confirming its potential to sustain reliable uplink communication in future sub-THz bands.
It should be noted that the lower SE observed at higher frequencies does not imply degraded communication capability.
The actual uplink throughput depends jointly on SE and available bandwidth.
Although higher frequencies suffer from stronger attenuation, they offer much wider bandwidths, which can ultimately lead to higher overall data rates.

\subsection{Robustness to Blockage}


In this subsection, we evaluate the uplink SE performance of {\SysName} and EdgeAnt under various blockage conditions to analyze the impact of user-induced signal obstruction.

\begin{figure}[!t]
  \centering
  \begin{minipage}[t]{0.48\columnwidth}
    \centering
    \includegraphics[width=\linewidth]{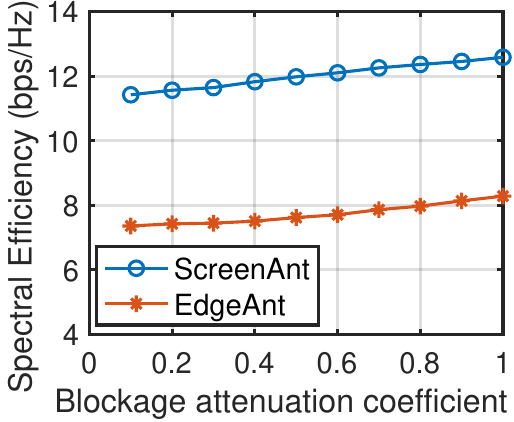}
    \captionof{figure}{SE versus the blockage attenuation coefficient $\beta$.} 
    \label{fig_blockage}
  \end{minipage}\hfill
  \begin{minipage}[t]{0.48\columnwidth}
    \centering
    \includegraphics[width=\linewidth]{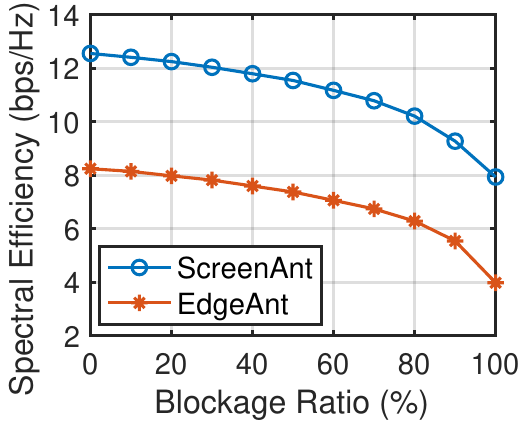}
    \captionof{figure}{SE versus the blockage ratio.} 
    \label{fig_ratio}
  \end{minipage}
  \vspace{-4mm}
\end{figure}

\textbf{Impact of the blockage attenuation coefficient.}
We first examine the effect of the blockage attenuation coefficient~$\beta$, which represents the fraction of the original signal power of an antenna element that remains when it is blocked.
Varying~$\beta$ models different blockage severities under the same blocked area---a smaller~$\beta$ corresponds to a closer (more severe) blockage, while a larger~$\beta$ corresponds to a farther (weaker) one.
The evaluation results are shown in Fig.~\ref{fig_blockage}. We observe that the achievable SE of both {\SysName} and EdgeAnt increases with~$\beta$, as a larger~$\beta$ implies weaker attenuation through the blockage.
Nevertheless, {\SysName} consistently outperforms EdgeAnt across all attenuation levels.

\textbf{Impact of the blockage ratio.}
We next evaluate the impact of the blockage ratio, which represents the proportion of antenna elements affected by user-induced obstruction.
As shown in Fig.~\ref{fig_ratio}, the achievable SE of both systems decreases with an increasing blockage ratio, as a larger portion of the array becomes blocked and contributes less to the overall radiation.
However, {\SysName} maintains a significantly higher SE than EdgeAnt across all blockage ratios.

It should be noted that, although {\SysName} and EdgeAnt are simulated under identical blockage parameters for fair comparison, their actual blockage behaviors differ in practice.
Antenna elements near the user’s hand experience stronger attenuation, and typical gripping postures result in higher blockage ratios for edge-mounted antennas.
During normal device usage, the user’s palm often covers the chassis area, leading to persistent and deep signal loss that cannot be easily mitigated.
In contrast, for {\SysName}, blockage mainly occurs when fingers touch the display surface.
Such contact is transient and spatially limited, resulting in dynamic but less severe attenuation.
Therefore, the real-world performance gap between {\SysName} and EdgeAnt is expected to be even larger than the simulated results suggest.

\begin{figure}[!t]
    \centering
    \includegraphics[width=0.99\columnwidth]{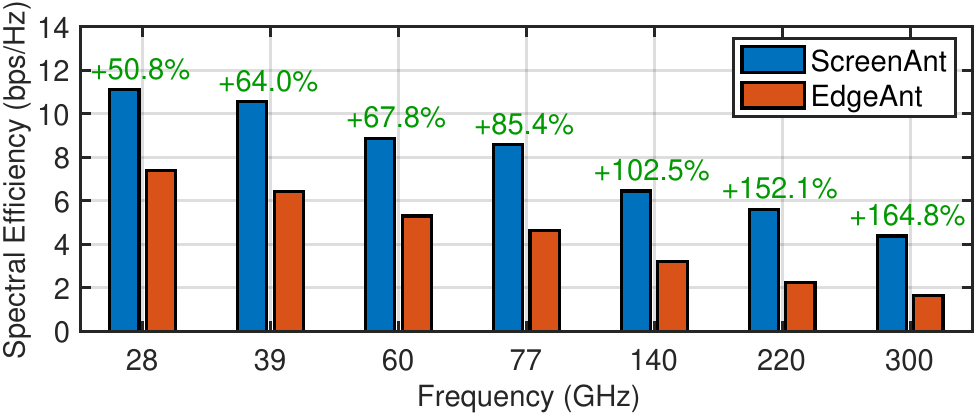}
    \captionof{figure}{SE versus carrier frequency $f_0$ (with blockage).} 
    \label{fig_frequency_blk}
    \vspace{-3mm}
\end{figure}

\textbf{Impact of carrier frequency.}
Finally, we evaluate the effect of the carrier frequency on the uplink SE under the blockage condition.
As shown in Fig.~\ref{fig_frequency_blk}, the achievable SE of both {\SysName} and EdgeAnt decreases as the carrier frequency increases from 28 GHz to 300 GHz, due to the stronger~path loss and more severe attenuation caused by blockage at higher carrier frequencies. The results are similar to those obtained in the no-blockage scenario shown in Fig.~\ref{fig_frequency}. 
However, under the existence of blockage, {\SysName} demonstrates much higher robustness compared to EdgeAnt, and the relative SE advantage of our {\SysName} over EdgeAnt further increases with the carrier frequency, from approximately 51\% at 28 GHz to about 165\% at 300 GHz.
This observation indicates that under practical blockage scenarios, high-frequency communication becomes increasingly challenging for EdgeAnt, whereas {\SysName} can maintain stable and reliable uplink performance.
Moreover, at higher frequencies, the smaller wavelength results in smaller antenna elements and shorter inter-element spacing, leading to a more compact on-screen array that is inherently less susceptible to hand blockage.
In contrast, edge-mounted antennas cannot avoid severe signal degradation caused by hand grip, even when their physical size is reduced, suggesting that the real-world advantage of {\SysName} would be greater than the simulated results indicate.

\section{Conclusion}
This paper presented \SysName, a novel on-screen antenna array architecture for future 6G user equipment (UE), where a fully transparent uniform planar array is seamlessly integrated onto the UE's screen.
A unified analytical and optimization framework was developed to characterize its uplink spectral efficiency (SE) under various parameters and blockage conditions.
The results show that, compared with conventional edge-mounted antennas, the proposed \SysName achieves over 50\% higher SE at 28 GHz in both conditions, with and without blockage, and achieves more than~150\%~gain~at~300~GHz.
Besides, \SysName shows superior robustness against hand-induced blockage. 
The results show the potential of \SysName as a promising technique to be deployed on future 6G UEs. 

\footnotesize 
\bibliographystyle{IEEEtran}
\bibliography{references}

\end{document}